\documentstyle{lamuphys}

\def\zgal{{${\rm z}_{\rm gal}$}}
\def\zabs{{${\rm z}_{\rm abs}$}}
\def\zform{{${\rm z}_{\rm form}$}}
\def\Boisse{Boiss\'e}

\def\H0#1{{H$_0$ = #1 km~s$^{-1}$ Mpc$^{-1}$}}
\input epsf
\makeatletter
\let\chapter\hid@chapter
\makeatother
\begin{document}
\pagenumbering{arabic}

\title{Implications for optical identifications of QSO absorption
systems from galaxy evolution models}

\author{Ulrich\,Lindner, 
Uta\, Fritze -- von Alvensleben, and
Klaus J.\,Fricke}

\institute{Universit\"ats-Sternwarte, Geismarlandstra{\ss}e 11,
D-37083 G\"ottingen, Germany}

\maketitle

\begin{abstract}
We have made an attempt to compile all currently available 
data on optically identified QSO absorber systems
(Lindner {\it et al.} 1996) to establish the status quo of 
absorber galaxy data as a basis for the investigation 
of galaxy evolution.
We present a first comparison with results from our galaxy
evolutionary synthesis models to demonstrate the potential
power of this kind of approach
and to guide future observations to identify absorber galaxies.
\end{abstract}

\section{Introduction and preliminary results}
Our chemical and spectral synthesis model 
(Fritze -- von Alvensleben {\it et al.} 1994) describes the 
evolution of various types of galaxies (E to Sd) with appropriate 
star formation histories and supplies us with time dependent 
values of luminosities from UV to NIR, colors and metallicities 
of model galaxies. Adopting any cosmological model characterized 
by $H_0$, $\Omega_0$, $\Lambda_0$ and the redshift of galaxy 
formation z$_{\small form}$, these results can be transformed 
into redshift dependent quantities. Apparent R 
magnitudes as a function of redshift z calculated with this 
model are plotted in Fig.~1.

Pioneering work in the optical identification of QSO
absorption systems was done by Bergeron \& \Boisse\ (1991) and
up to now, there are more than a dozen of publications reporting
on photometric data, equivalent widths and impact parameters
for absorbing galaxies. All available data on
apparent R magnitudes are plotted in Fig.~1. Different symbols
correspond to different authors. Absorbing galaxies with
spectroscopically confirmed redshifts (i.e. \zgal\ = \zabs)
are marked by filled symbols, whereas
open symbols indicate absorber candidates. 

Accounting for the luminosity ranges from the luminosity functions
of the various galaxy types (cf. $2 \sigma_R$ bars in the lower 
right corner of Fig.~1) we can state that 
virtually all observational data points fall between the curves 
for E-- and Sd--models and, accordingly, we can establish global 
agreement between our galaxy evolution models and observational 
data up to $z \approx 2$.

Most of the absorber galaxies appear to be early through 
intermediate type spirals (Sa--Sbc) but many ellipticals 
and some late type spirals seem to be present, too.
The presence of intermediate and late type galaxies among QSO 
absorbers would imply that a considerable fraction 
of these galaxies do have extended gaseous halos metal rich enough 
to cause detectable absorption.

\begin{figure}
\epsfysize=14.5cm
\vskip -6.5truecm
{\epsffile{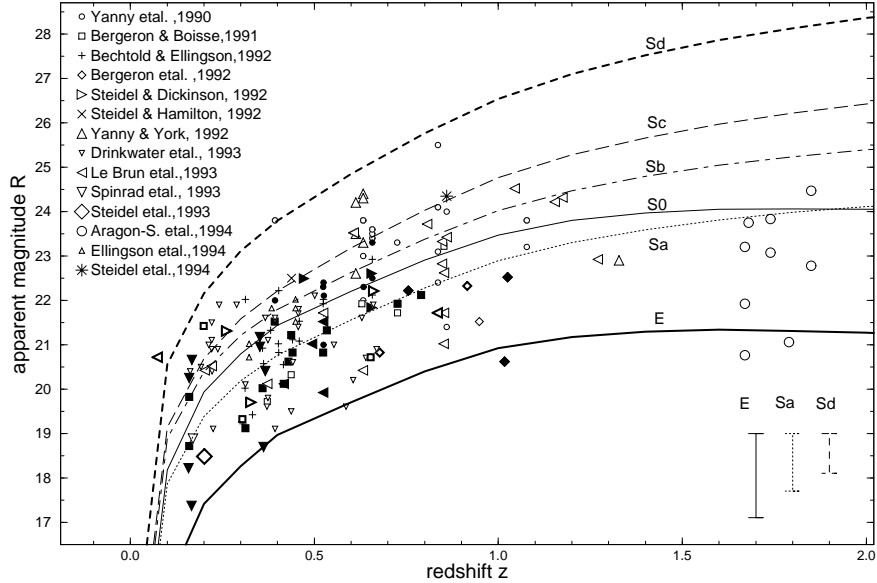}}
\vskip -0.3truecm
\caption{Apparent R magnitude as a function of redshift for observational data 
and results from our galaxy evolution model using \zform\ $= 5$ and 
cosmological parameters \H0{50}, $\Omega_0 = 1$, and $\Lambda_0 = 0$.}
\end{figure}

Varying the cosmological parameters we find that e.g. model
galaxies for \H0{50} and $\Omega_0 = 0.1$ are much too 
faint as compared to observations leading us
to exclude this compination of cosmological parameters.

\section{Outlook and VLT perspective}
Unfortunately, data are very inhomogeneous at present. 
Measurements in different passbands, i.e. colors, from
present day instruments are needed.
FORS attached to the VLT will
provide deeper imaging ($24 < R < 26$) to detect
intermediate and late type spiral galaxies in the
redshift range $1 < z < 2$ and the MOS unit of FORS
will allow to spectroscopically confirm absorber
candidates in the magnitude range $22 < R < 24$ (cf. Fig.~1).

\bigskip\noindent
{\bf Acknowledgement}

\noindent
This work was supported by Verbundforschung
Astronomie/Astrophysik through BMFT grant FKZ~50~0R~90045
%
%

\end{document}